\documentclass{PoS}
\title     {$I=2$ Two-Pion Wave Functions with Non-zero Total Momentum}
\ShortTitle{$I=2$ Two-Pion Wave Functions with Non-zero Total Momentum}
\author{\speaker{Kiyoshi Sasaki} \\
  Center for Computational Science, University of Tsukuba, \\
  Tsukuba, Ibaraki 305-8577, Japan \\
  E-mail: \email{ksasaki@ccs.tsukuba.ac.jp}
}
\author{Naruhito Ishizuka \\
  Graduate School of Pure and Applied Sciences, University of Tsukuba, \\
  Tsukuba, Ibaraki 305-8571, Japan \\
  and\\
  Center for Computational Science, University of Tsukuba, \\
  Tsukuba, Ibaraki 305-8577, Japan \\
  E-mail: \email{ishizuka@het.ph.tsukuba.ac.jp}
}
\abstract{
We calculate the two-pion wave function 
for the $I=2$ $S$-wave two-pion system
with a finite scattering momentum
and estimate the interaction range between two pions. 
It allows us to examine the validity of the necessary condition
for the finite-volume method for the scattering phase shift.
A calculation is carried out 
with a plaquette gauge action for gluons 
and a clover-improved Wilson action for quarks 
at $1/a=1.63\ {\rm GeV}$ on $32^3\times 120$ lattice 
in the quenched approximation. 
We conclude that 
the necessary condition is satisfied 
within statistical errors for the lattice size $L\ge 32$,
when the quark mass is in the range 
$m_\pi^2=0.176 - 0.345\ {\rm GeV}^2$ and 
the scattering momentum in $k^2 < 0.026\ {\rm GeV}^2$.
We also find that 
the energy dependence of the interaction range
is small and it takes $1.2-1.7\ {\rm fm}$
for our simulation parameters.
We obtain the phase shift from the two-pion wave function
with a smaller statistical error
than that from 
the conventional analysis with the two-pion time correlator. 
}
\FullConference{
  The XXV International Symposium on Lattice Field Theory \\
  July 30 - August 4 2007 \\
  Regensburg, Germany
}
\begin{document}
%
%
\section  {Introduction}
\label{sec:Introduction}
The scattering phase shift is an important quantity 
for understanding a dynamical aspect of hadrons.
For the $I=2$ $S$-wave two-pion system, which is the simplest case, 
the phase shift has been calculated 
in Refs.~\cite{
RelatedWorks1, 
RelatedWorks2,
RelatedWorks3,
RelatedWorks4}.
The calculations employed the finite-volume method, 
in which the phase shift is related to the energy on a finite volume.
It has been proposed by L\"uscher~\cite{FSM}
and extended for the non-zero momentum system
by Rummukainen and Gottlieb~\cite{Rummukainen:1995vs}. 

The derivation of L\"uscher's and Rummukainen-Gottlieb's formula
assumes the condition $R<L/2$
for the two-pion interaction range $R$ and the lattice size $L$, 
so that the boundary condition does not distort 
the shape of the two-pion interaction. 
The CP-PACS collaboration calculated the two-pion wave function  
for the ground state of the $I=2$ $S$-wave two-pion system 
and estimated the interaction range $R$  
from the wave function~\cite{Aoki:2005uf}.
In their case the scattering momentum is highly small, $k \sim 0$, 
thus their work is an examination of the necessary condition 
for the calculation of the scattering length on the lattice.

In present work
we extend their work for the scattering length  
(the scattering momentum $k\sim0$) 
to that for the scattering phase shift ($k\not= 0$). 
For this aim
we consider the ground state of the system 
having a non-zero total momentum
${\bf P}=(2\pi / L){\bf e}_x$
in a $L^3$ box satisfying the periodic boundary condition.
All calculations of this work has been done on VPP5000/80 at the 
Academic Computing and Communications Center of University of Tsukuba. 
%
%
\section  {Finite size formula}
\label{sec:Finite size formula}
In this section 
we briefly review the finite-volume method 
presented by Rummukainen and Gottlieb 
in Ref.~\cite{Rummukainen:1995vs},
with emphasis on the role of the condition
for the interaction range.
The formula has also been derived from another approaches 
in Ref.~\cite{Kim:2005gf} and \cite{Christ:2005gi}.
We follow, however, from the original derivation 
in Ref.~\cite{Rummukainen:1995vs}. 

The two-pion wave function on a finite periodic box 
of volume $L^3$ is defined by
\begin{equation}
  \Phi({\bf x},t)
  =  \langle 0 | \ \pi^{+}({\bf X} + {\bf x}/2, T + t/2 )
                 \ \pi^{+}({\bf X} - {\bf x}/2, T - t/2 ) 
     \ | \pi\pi;E,{\bf P}\rangle
             \cdot {\rm e}^{ -i {\bf P}\cdot{\bf X} }
             \cdot {\rm e}^{ E \cdot T              }
\ ,
\label{eq:eq_def_wf}
\end{equation}
where $|\pi\pi;E,{\bf P}\rangle$ 
is an energy eigenstate of the two-pion system 
with the energy $E$ and the total momentum ${\bf P}$.
$\pi^{+}({\bf x},t)$ is
an interpolating operator for $\pi^{+}$ at $({\bf x},t)$.
The two exponential factors are introduced 
to remove the trivial exponential factors 
for the center of mass coordinate ${\bf X}$ and $T$.

In order to relate the wave function to the scattering phase shift,
we transform the wave function $\Phi({\bf x},t)$ in (\ref{eq:eq_def_wf})
to that in the center of mass frame 
$\Phi_{\rm CM}({\bf x},t)$ 
by the Lorentz transformation.
Here we assume 
that the two-pion interaction range $R$ 
is smaller than one half the lattice extent, 
{\it i.e.} there exists the region $R < |{\bf x}| < L/2$, 
where the two pions behave as free particles.
In this region 
the wave function satisfies 
the following two equations~\cite{Rummukainen:1995vs}.
\begin{equation}
  ( \nabla^2 + k^2 )\ \Phi_{\rm CM}({\bf x},t) = 0 \quad , \quad
  \frac{\partial}{\partial t} \ \Phi_{\rm CM}({\bf x},t) = 0
\ ,
\label{eqn:HlEq}
\end{equation}
where $k$ is the scattering momentum 
related to the invariant mass by
$\sqrt{s}=\sqrt{E^2-P^2}=2\sqrt{m_\pi^2+k^2}$.
$\Phi_{\rm CM}({\bf x},t)$ also satisfies 
the boundary condition,
\begin{equation}
  \Phi_{\rm CM}({\bf x},t)
  =  (-1)^{ (L/2\pi) {\bf P}\cdot{\bf m} }
     \cdot
     \Phi_{\rm CM}( {\bf x} + L \hat{\gamma}[{\bf m}], t )
               \qquad \mbox{ for \ ${\bf m} \in \mathbb{Z}^3$ }
\ ,
\label{eqn:bc}
\end{equation}
where
$\hat{\gamma}$ is the vector operation 
$\hat{\gamma}[{\bf x}]=
\gamma {\bf x}_{\parallel} + {\bf x}_{\perp}$
with the Lorentz boost factor $\gamma = E/\sqrt{s}$,
${\bf x}_{\parallel} = {\bf P}({\bf P}\cdot{\bf x})/P^2$ and 
${\bf x}_{\perp}     = {\bf x} - {\bf x}_{\parallel}$.

The solution of the (\ref{eqn:HlEq}) 
under the condition (\ref{eqn:bc}) can be given by 
\begin{eqnarray}
      \Phi_{\rm CM}({\bf x},t)
  &=&
     \frac{1}{\gamma L^3}
     \sum_{ {\bf p}\in\Gamma }
        \ \frac{1}{p^2-k^2}
          \cdot
          \mbox{e}^{ i {\bf p}\cdot{\bf x} }
  \ , \ 
  \Bigl(\ \Gamma = \left\{ 
          \ {\bf p}\ | 
          \ {\bf p} = (2\pi/L) \cdot \hat {\gamma}^{-1}[{\bf n}]  +  {\bf P}/2 , 
          \ {\bf n}\in \mathbb{Z}^3 \right\}
 \Bigr)
\label{eqn:sl1}
\\
  &=&
    \frac{1}{\gamma L^3}
    \sum_{ {\bf p} \in\Gamma}\ \frac{1}{ p^2 - k^2 }
                   \cdot j_0 (kx)
  + \frac{k}{4\pi} \cdot n_0 (kx)
  + \sum_{lm} C_{lm}(k) \cdot Y_{l}^{m}(\Omega) j_l(kx)
\ ,
\label{eqn:sl2}
\end{eqnarray}
up to overall constant,
where $j_l(kx)$ is the spherical Bessel 
and $n_l(kx)$ is the spherical Neumann function.
$C_{lm}(k)$ is some constant depending on the scattering momentum $k$.
The first and second terms of (\ref{eqn:sl2})
consist of the $S$-wave component
and those coefficients
give the $S$-wave phase shift $\delta(k)$,
\begin{equation}
  \frac{1}{ \tan \delta_0 (k)}
  =
  \frac{4\pi}{k}
  \cdot
  \frac{1}{\gamma L^3 }
  \sum_{ {\bf p} \in\Gamma}\ \frac{1}{ p^2 - k^2 }
\ .
\label{eqn:RG}
\end{equation}
This is the Rummukainen-Gottlieb formula~\cite{Rummukainen:1995vs}. 
%
%
\section  {Details of simulation}
\label{sec:Details_of_simulation}
In the present work
we consider the ground state of the system with the total momentum  
${\bf P}={\bf 0}$ and ${\bf P}=(2\pi/L){\bf e}_x$.
We can obtain the scattering length 
from the energy of the system with ${\bf P}={\bf 0}$
and the phase shift 
from that with ${\bf P}=(2\pi/L){\bf e}_x$
through the Rummukainen-Gottlieb formula (\ref{eqn:RG}).

In order to calculate the wave function
we consider the correlator,
\begin{equation}
  F({\bf x},\tau)
    = \langle 0| \           \Omega ({\bf x},\tau  ) 
                 \ \overline{\Omega}({\bf P},\tau_s) \ |0 \rangle
\ .
\label{eqn:four-point-func}
\end{equation}
The operator 
$\Omega({\bf x},\tau)$ is defined by 
\begin{equation}
  \Omega({\bf x},\tau)
     = \sum_{ \hat{R} }
       \sum_{ {\bf X} }
               \ {\rm e}^{ i {\bf P} \cdot {\bf X} }
               \ \pi^+ ( {\bf X} + \hat{R}[{\bf x}] , \tau )
               \ \pi^+ ( {\bf X}                    , \tau )
\ .
\label{eqn:def_Omega}
\end{equation}
The vector operation $\hat{R}$ represents an element of 
the cubic      group ($O_h   $) for ${\bf P}={\bf 0}$ and
the tetragonal group ($D_{4h}$) for ${\bf P}=(2\pi/L){\bf e}_x$.
The summation over $\hat{R}$
projects out ${\bf A_1^+}$ representation of these groups,
which equals to the $S$-wave state
ignoring the effects from higher angular momentum $l\ge 2$.

The operator $\overline{\Omega}({\bf P},\tau_s)$ 
in (\ref{eqn:four-point-func}) is defined by
\begin{equation}
  \overline{\Omega}({\bf P},\tau_s)
    = \frac{ 1 }{ N_R }
      \sum_{ j=1 }^{ N_R }
         \left[  \pi^+ ( {\bf P}, \tau_s ; {\xi }_j )
                 \pi^+ ( {\bf 0}, \tau_s ; {\eta}_j ) \right]^\dagger
\ , 
\label{eqn:def_Omega_source}
\end{equation}
where 
\begin{equation}
 \pi^+( {\bf P}, \tau_s ; {\xi }_j )
    = 
      \left[  \sum_{ {\bf x} }
                {\rm e}^{ i {\bf P}\cdot{\bf x} } 
                \bar{u}( {\bf x}, \tau_s ) 
                \xi_j^\dagger ({\bf x} )
      \right]
           \gamma_5
      \left[  \sum_{ {\bf y} }
                d( {\bf y}, \tau_s ) \xi_j ({\bf y})
      \right]
\ .
\end{equation}
The operator  ${\pi}^{+}({\bf P}, \tau_s ; {\eta}_j  )$
is defined as ${\pi}^{+}({\bf P}, \tau_s ; {\xi }_j  )$
by changing $\xi_j({\bf x})$ to $\eta_j({\bf x})$.
The functions $\xi_j({\bf x})$ and $\eta_j({\bf x})$ are
$U(1)$ noise whose property is
\begin{equation}
  \lim_{ N_R \to \infty } 
      \frac{1}{N_R} \sum_{j=1}^{N_R} 
             \xi_j^* ({\bf x}) \xi_j ({\bf y}) 
      = \delta^3 ( {\bf x} - {\bf y} ) 
\ .
\end{equation}
In the present work
we take $N_R=2$
in (\ref{eqn:def_Omega_source}).

In large $\tau$ region,
we can obtain the wave function 
for the ground state in (\ref{eq:eq_def_wf}) by
$\Phi({\bf x},0) =  F({\bf x},\tau) / F({\bf x}_0,\tau)$
introducing the reference position ${\bf x}_0$.
In the present work
we set $\tau_s=20$, $\tau-\tau_s=40$ and ${\bf x}_0=(7,5,2)$.
In the estimation of the interaction range,
we analyze the wave function in the center of mass frame
$\Phi_{\rm CM}({\bf x},t)$ at $t=0$.
It is related to $\Phi({\bf x},t)$ by 
$\Phi_{\rm CM} (\hat{\gamma}[{\bf x}], 0 ) = 
 \Phi          (             {\bf x} , 0 )$
outside of the interacting region ($|{\bf x}|>R$)
from the second equation in (\ref{eqn:HlEq}).

Gauge configurations are generated in a quenched approximation 
with a plaquette gauge action at $\beta=5.9$ 
on a $32^3\times 120$ lattice. 
The physical quantities are measured every 200 sweeps 
independently for each quark mass parameter. 
A clover fermion action with $C_{SW}=1.364$ is used.
The quark propagators are imposed
to the Dirichlet boundary condition in the time direction 
and to the periodic boundary condition in the spatial one. 
The lattice cutoff is estimated as 
$1/a=1.63(5)\ {\rm GeV}$ ($a=0.121(3)$ fm)
from the $\rho$ meson mass.
Three quark masses are chosen
to give $m_\pi^2=0.176$, $0.238$ and $0.345\ {\rm GeV^2}$. 
The numbers of configurations are $400$, $212$ and $212$ 
for each quark masses.
%
%
\section{ Results of interaction range }
\begin{figure}[t]
\begin{center}
\vspace{-4pt}
\begin{tabular}{ll}
\multicolumn{2}{l}{(a) ${\bf P}={\bf 0}$} \\
\includegraphics[width=55mm]{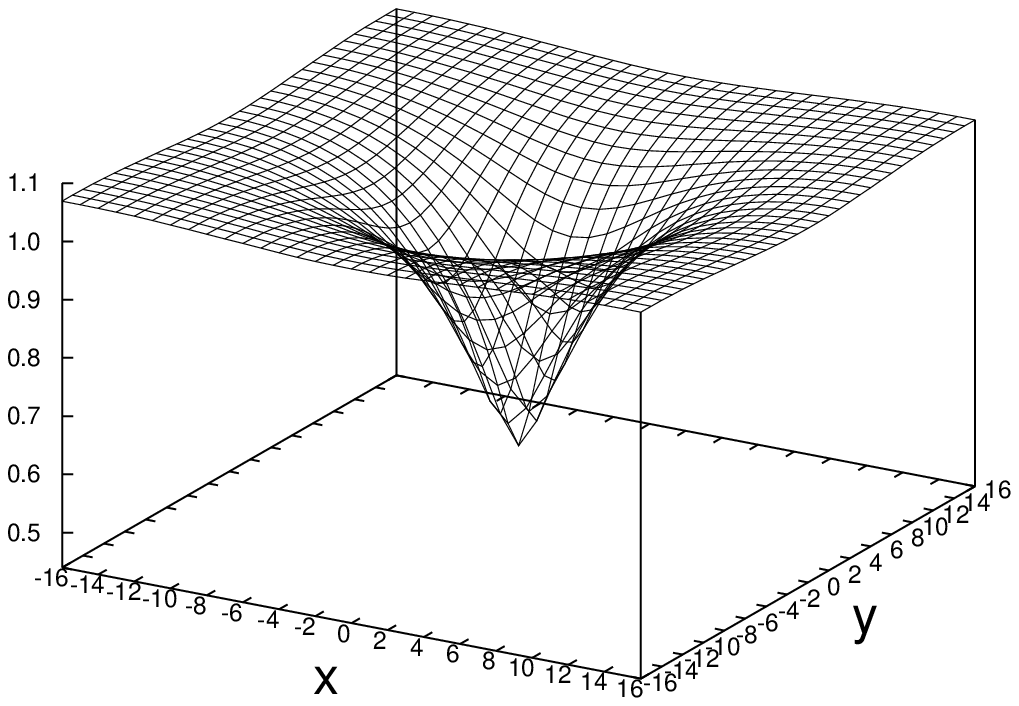}
\includegraphics[width=55mm]{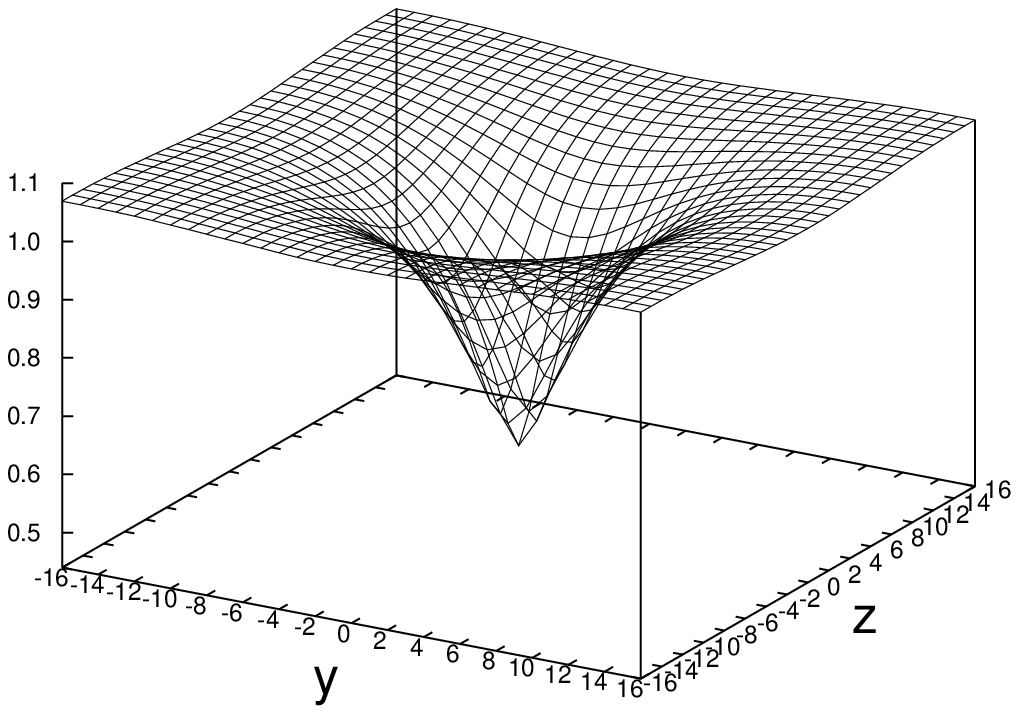} & 
\\
\multicolumn{2}{l}{(b) ${\bf P}=(2\pi/L){\bf e}_x$}  \\
\includegraphics[width=55mm]{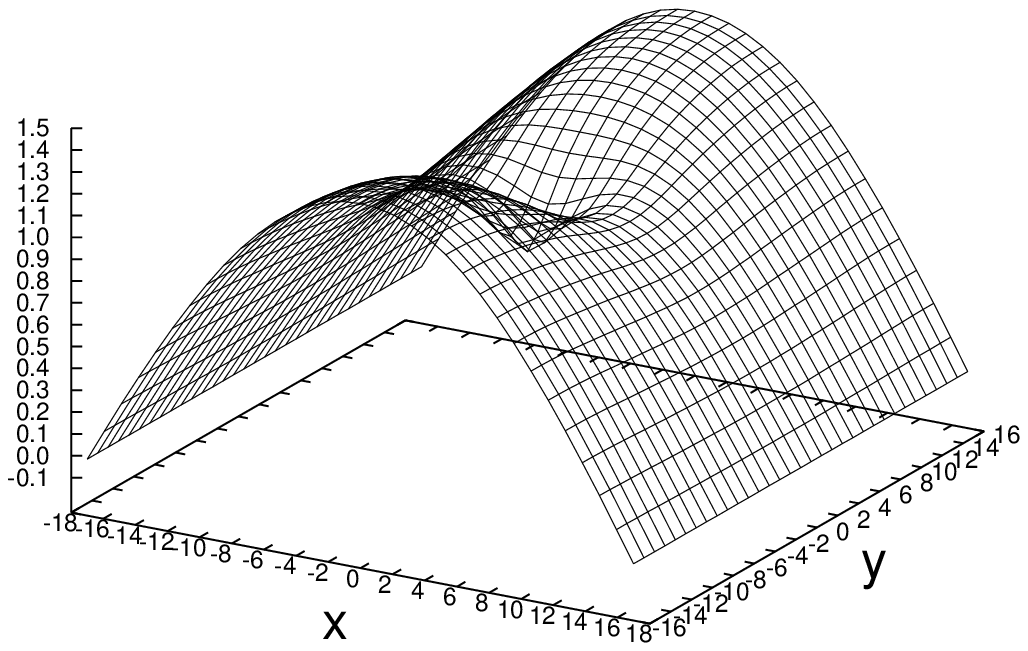}
\includegraphics[width=55mm]{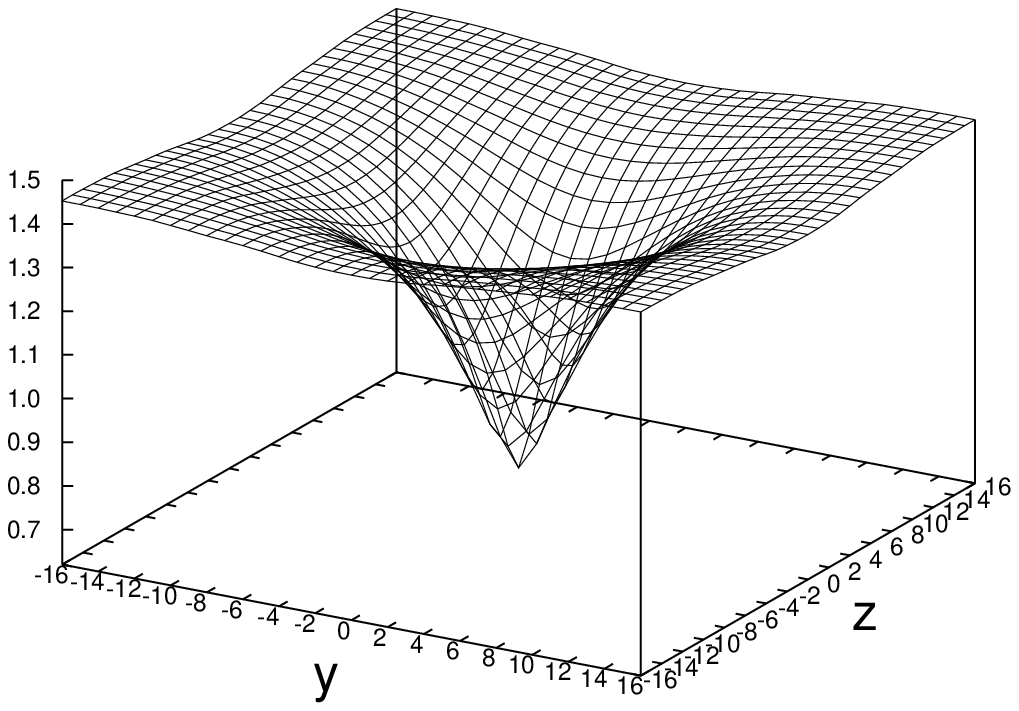} & 
\end{tabular}
\caption{
The two-pion wave functions $\Phi_{\rm CM}({\bf x})$
at $m_\pi^2=0.176$ GeV${}^2$
for ${\bf P}={\bf 0}$ and ${\bf P}=(2\pi/L){\bf e}_x$.
The left and right panels for each momentum show
the wave function on $xy$- and $yz$-plane. 
}
\label{fig:wave-function-3d_1}
\end{center}
\end{figure}

In Fig.~\ref{fig:wave-function-3d_1}, 
we show the two-pion wave functions $\Phi_{\rm CM}({\bf x})$
at $m_\pi^2=0.176\ {\rm GeV}^2$
for ${\bf P}={\bf 0}$ 
and for ${\bf P}=(2\pi/L){\bf e}_x$.
The left and right panels for each momentum show
the wave function on $xy$- and $yz$-plane. 
We find a very clear signal.
%

\begin{figure}[t]
\begin{center}
\begin{tabular}{ll}
\multicolumn{2}{l}{(a) ${\bf P}={\bf 0}$} \\
\includegraphics[width=55mm]{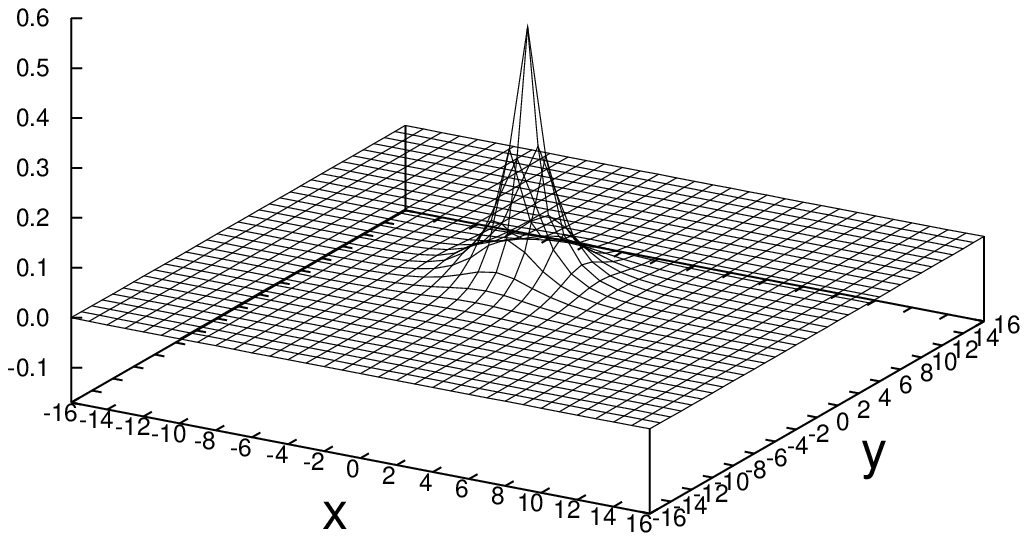} &
\includegraphics[width=55mm]{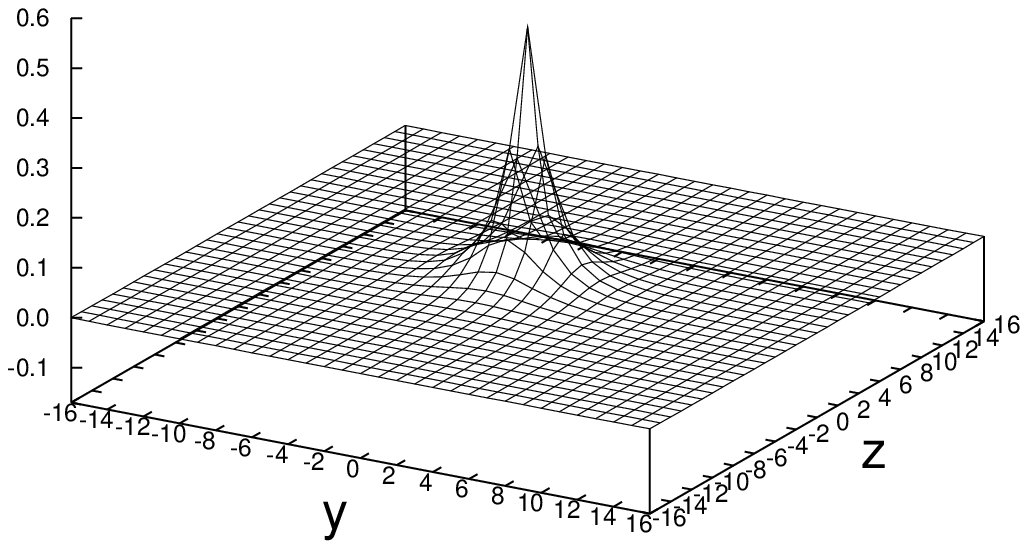}
\\
\multicolumn{2}{l}{(b) ${\bf P}=(2\pi/L){\bf e}_x$} \\
\includegraphics[width=55mm]{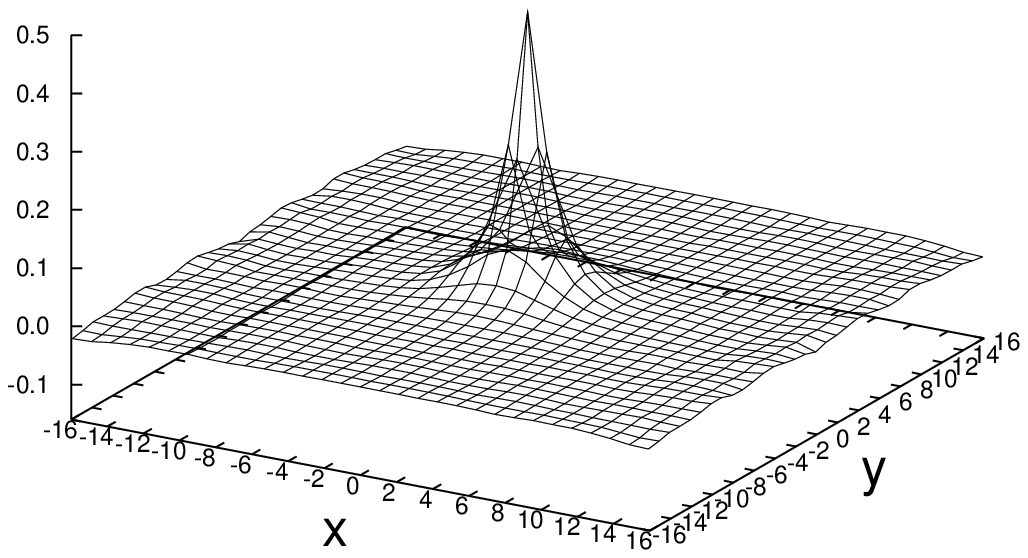} &
\includegraphics[width=55mm]{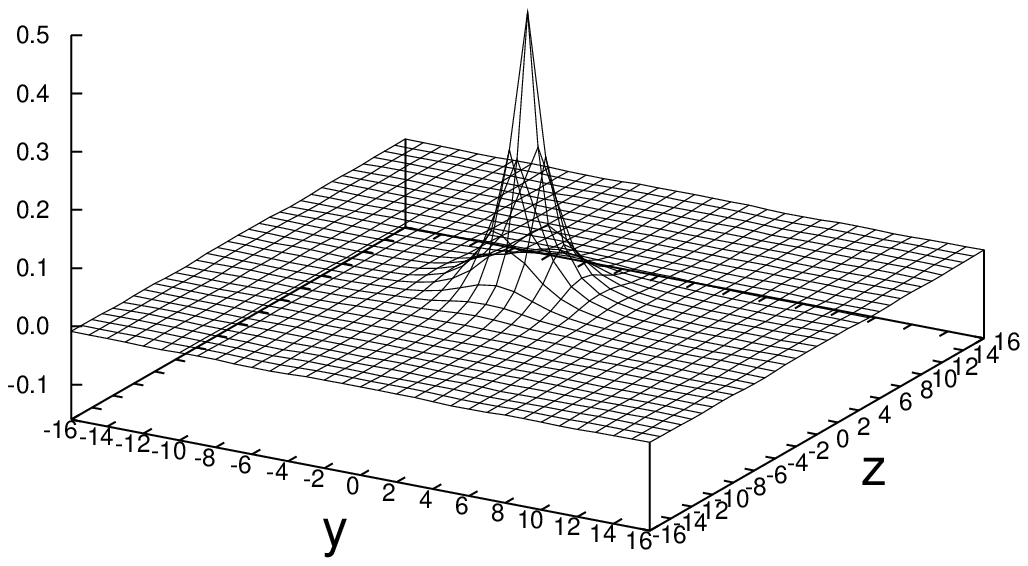}
\end{tabular}
\caption{
$V({\bf x})$ at $m_\pi^2=0.176$ GeV${}^2$
for ${\bf P}={\bf 0}$ and ${\bf P}=(2\pi/L){\bf e}_x$.
The left and right panels for each momentum show
the $V({\bf x})$ on $xy$- and $yz$-plane. 
}
\label{fig:Vx-3d}
\end{center}
\end{figure}

We now consider the two-pion interaction from the ratio,
$V({\bf x})=\nabla^2 \Phi_{\rm CM} ({\bf x}) / \Phi_{\rm CM}({\bf x})$.
Away from the two-pion interaction range,
{\it i.e.} $|{\bf x}|>R$, we expect that 
$V({\bf x})$ is independent of ${\bf x}$ and equals to $-k^2$
from (\ref{eqn:HlEq}).
In Fig.~\ref{fig:Vx-3d}, 
$V({\bf x})$ for the same parameters 
as for Fig.~\ref{fig:wave-function-3d_1}
are plotted.
We find a very clear signal 
and $V({\bf x})$ seems to be constant for $|{\bf x}|>10$. 
We observe a strong repulsive interaction at the origin 
consistent with the negative phase shift 
of the $I=2$ two-pion system. 
%

In order to estimate the interaction range $R$,
we consider, 
\begin{equation}
  U({\bf x}) = \nabla^2 \Phi_{\rm CM}({\bf x})
                      / \Phi_{\rm CM}({\bf x}) + k^2
\ , 
\end{equation}
where $k^2$ is obtained 
from the two-pion time correlator. 
According to Ref.~\cite{Aoki:2005uf}, 
we employ the operational definition of the interaction range $R$
as the scale where 
$U({\bf x})$ is sufficiently small compared to the statistical error. 
Strictly speaking, 
even if $|{\bf x}|$ takes a large value,
$U({\bf x})$ does not vanish 
and has a finite tail.
However, the systematic error 
for the final results of the phase shift 
due to the existence of the tail
is buried into the statistical error in this definition. 

\begin{figure}[t]
\begin{center}
\begin{tabular}{l}
\multicolumn{1}{l}{(a) ${\bf P}={\bf 0}$}                    \\
\includegraphics[height=45mm]{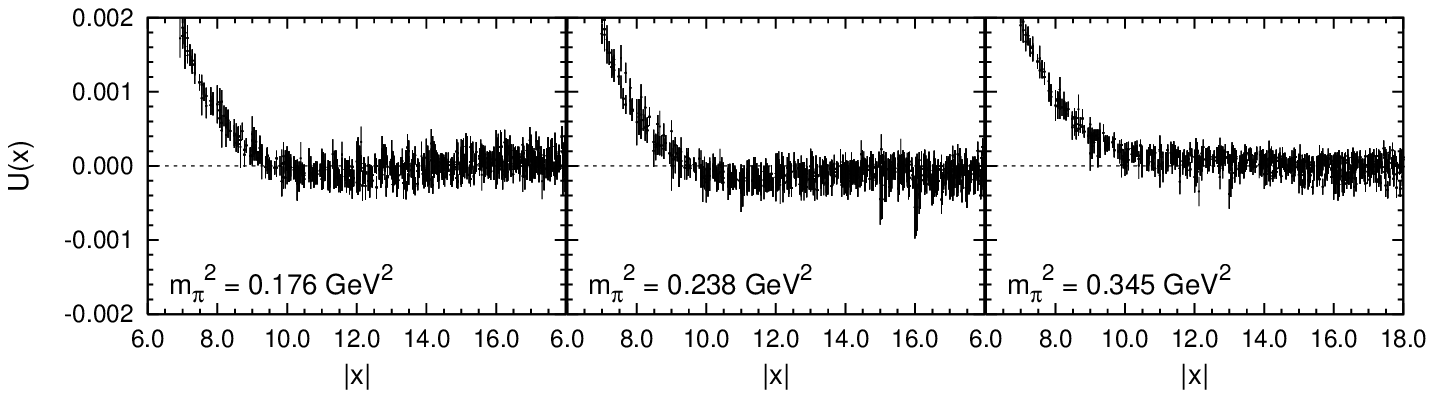}  \\
\multicolumn{1}{l}{(b) ${\bf P}=(2\pi/L){\bf e}_x$}    \\
\includegraphics[height=45mm]{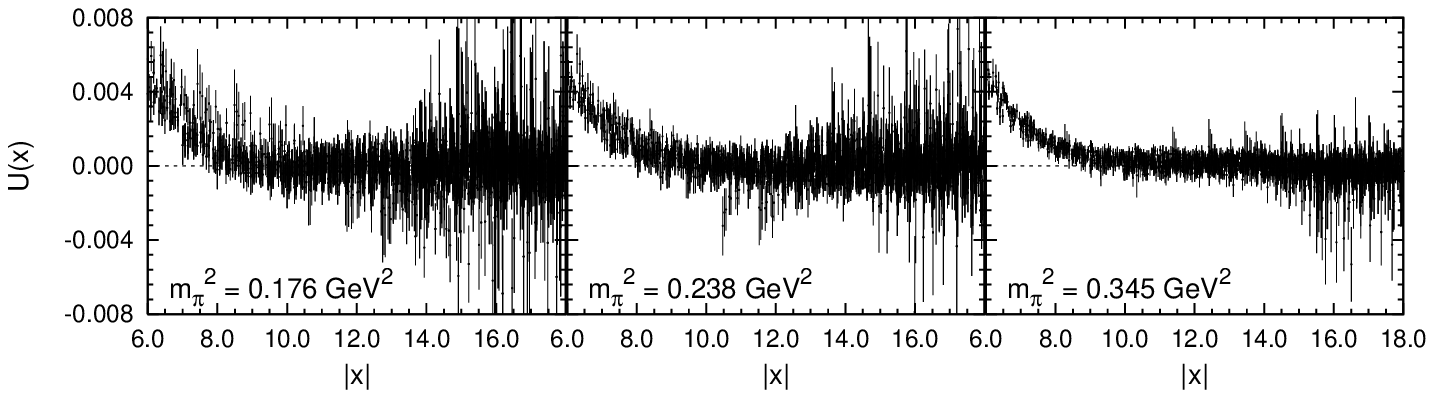}
\end{tabular}
\caption{
$U({\bf x})$ as a function of $|{\bf x}|$
for ${\bf P}={\bf 0}$ and ${\bf P}=(2\pi/L){\bf e}_x$
for several quark masses.
}
\label{fig:edp-of-int-range}
\end{center}
\end{figure}

In Fig.~\ref{fig:edp-of-int-range}, 
we show $U({\bf x})$ as a function of $|{\bf x}|$
for ${\bf P}={\bf 0}$ and ${\bf P}=(2\pi/L){\bf e}_x$
at the three quark masses.
We find that
the interaction range $R$ takes 
\begin{eqnarray}
&&  
m_\pi^2\hspace{2pt}(\mbox{GeV}^2)\hspace{40pt}  
\hspace{22pt}  0.176    
\hspace{16pt}  0.238  
\hspace{16pt}  0.345  \cr
&&  
R\hspace{8pt}\mbox{for}\hspace{6pt}{\bf P}={\bf 0}\hspace{36pt}  
\hspace{22pt}   13.0
\hspace{22pt}   14.0
\hspace{22pt}   12.0  \cr
&&  
R\hspace{8pt}\mbox{for}\hspace{6pt}{\bf P}=(2\pi/L){\bf e}_x  
\hspace{22pt}   10.0
\hspace{22pt}   11.0
\hspace{22pt}   12.0
~ ~ .
\label{eqn:int-range}
\end{eqnarray}
It is at most $14.0$ ($1.69\ {\rm fm}$) 
and smaller than $L/2=16$.
Thus
the necessary condition 
for the Rummukainen-Gottlieb formula (\ref{eqn:RG})
is satisfied within statistical errors
at our simulation points.
%
%
%
\section{ Results of scattering phase shift }
\begin{figure}[t]
\begin{center}
\vspace{-11pt}
\includegraphics[height=52mm]{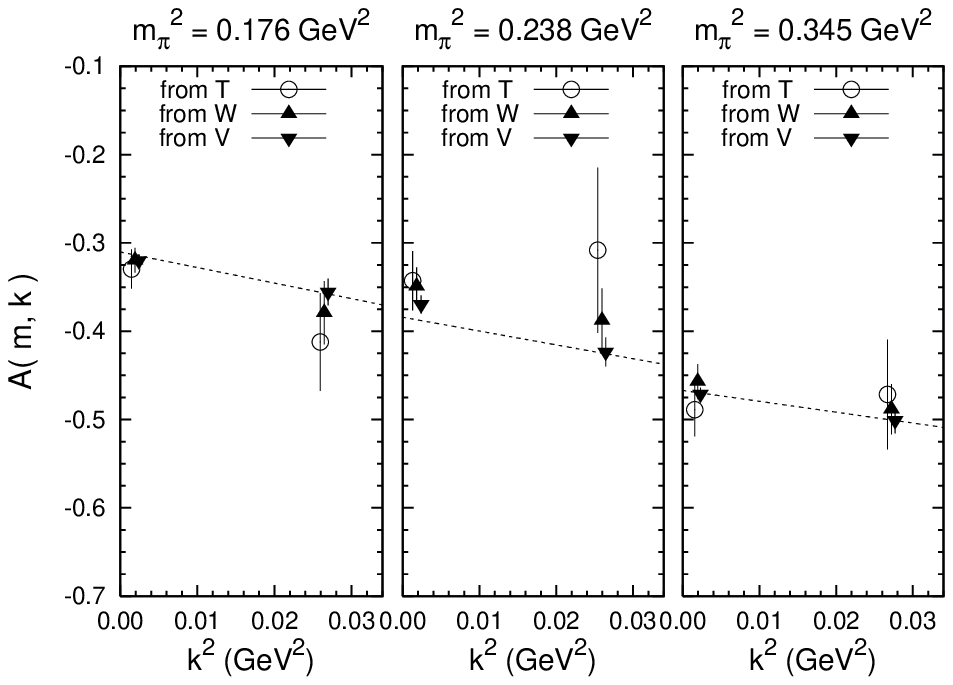}
\includegraphics[height=52mm]{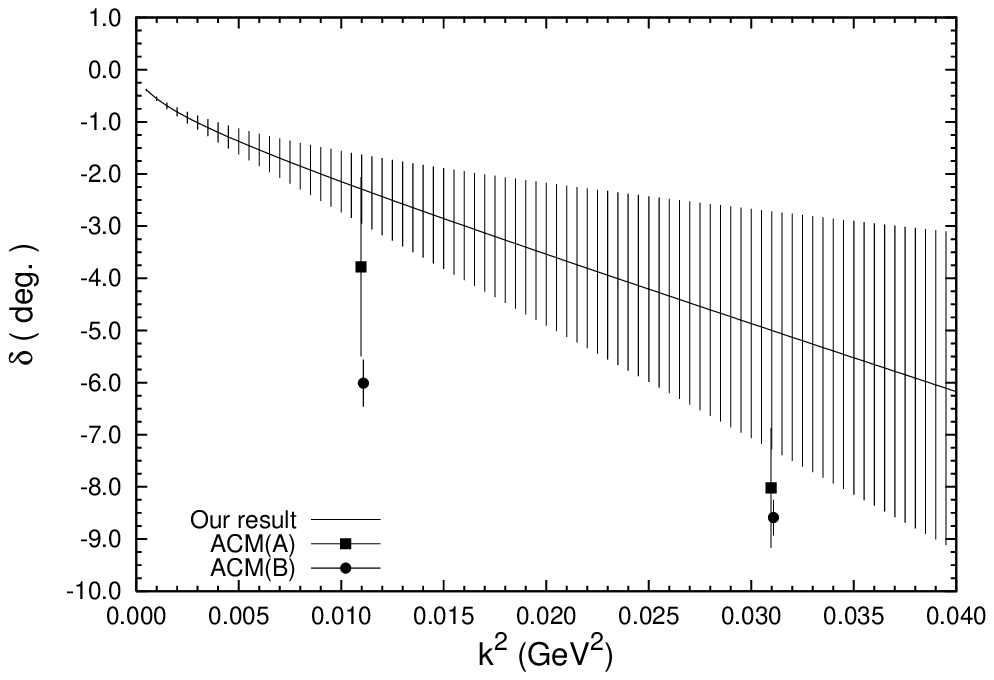}
\caption{
In the left panel
$A(m_\pi,k)$ given by the three methods
are shown as a function of $k^2$. 
The dotted line is the fit curve 
for the data given by the method (``from V'').
In the right panel
our final results of the scattering phase shift at the physical quark mass 
are plotted and compared with the experiment~\cite{Hoogland:1977kt}.
}
\label{fig:k2-amp-delta}
\end{center}
\end{figure}

We estimate the phase shift 
with the following three methods :
%
\renewcommand{\labelenumi}{\arabic{enumi}.} 
\begin{enumerate} 
\item \label{method_T} 
We extract the energy $E$ from the two-pion time correlator
and calculate the scattering momentum by 
$k^2 = ( E^2 - P^2 )/4 - m_\pi^2$. 
The phase shift is calculated 
by substituting $k^2$ into (\ref{eqn:RG}).
The results of the phase shift are shown 
in the left panel of Fig.~\ref{fig:k2-amp-delta} (labeled ``from T''),
where the scattering amplitudes,
\begin{equation}
  A( m_\pi, k )= \tan\delta_0(k) / k \cdot \sqrt{m_\pi^2+k^2}
\end{equation}
are plotted.
%
\item \label{method_W} 
We extract $k^2$ by
fitting the wave function $\Phi_{\rm CM}({\bf x})$ 
with the fitting function given in (\ref{eqn:sl1})
taking $k^2$ and an overall constant as the fitting parameters.
We choose the fitting range $|{\bf x}|>R$ 
with $R$ given in (\ref{eqn:int-range}). 
The results are plotted 
in the left panel of Fig.~\ref{fig:k2-amp-delta} (labeled ``from W'').
%
\item \label{method_V} 
$k^2$ is extracted 
by fitting $V({\bf x})$ to a constant
in the region $|{\bf x}|>R$ with $R$ given in (\ref{eqn:int-range}).
We show the results
in the left panel of Fig.~\ref{fig:k2-amp-delta} 
(labeled ``from V'').
\end{enumerate} 
%
As shown in the left panel of Fig.~\ref{fig:k2-amp-delta},
the results given by the three method 
are consistent within the statistical errors.
The data given by the method~\ref{method_V} 
(``from V'') provides the smallest statistical error, 
so the following analysis is performed with this data. 

In order to obtain the phase shift at the physical quark mass,  
we extrapolate our results with the fit form,
\begin{equation}
  A(m_\pi, k) =  A_{10}m_\pi^2     + A_{20}m_\pi^4
              +  A_{01}        k^2 + A_{11}m_\pi^2 k^2
\ .
\label{eqn:expansion-of-amplitude}
\end{equation}
The fit curves for this fitting are also plotted 
in the left panel of Fig.~\ref{fig:k2-amp-delta}.
As shown in the figure the fitting is carried out well.
Our final results of the phase shift 
at the physical quark mass
are shown in the right panel of Fig.\ref{fig:k2-amp-delta}
and compared with the experiment~\cite{Hoogland:1977kt}.
Our results are slightly larger than the experiment.
A possible origin of the discrepancy is finite lattice spacing effects. 
We must leave the confirmation of this 
to studies in the future. 
%
%
%
\section{Conclusion}
\label{sec:Conclusion}
In the present work, we have studied 
the $I=2$ two-pion wave functions 
with the scattering momentum $k^2\sim 0$ and $k^2\ne 0$. 
We have estimated the two-pion interaction ranges $R$ from those. 
It has been confirmed that 
the necessary condition for the Rummukainen-Gottlieb formula (\ref{eqn:RG}) 
satisfies within statistical errors in our parameters.
Moreover, we have estimated the scattering phase shift 
with the two-pion wave function. 
These methods provide a smaller statistical error 
than that from the conventional analysis with the two-pion time correlator. 
%
%

%
\end{document}